\documentclass{PoS}

\usepackage{isotope}

\newcommand{\n}[1]{\ensuremath{|\mathbf{#1}|}}

\graphicspath{{plot/}}

\title{Spectral functions of argon and calcium}

\ShortTitle{Spectral functions of argon and calcium}

\author{\speaker{Artur M. Ankowski}\\
        Institute of Theoretical Physics\\ University of Wroc{\l}aw\\ pl. Maksa Borna 9\\ 50-204 Wroc{\l}aw\\ Poland\\
        E-mail: \email{artank@ift.uni.wroc.pl}}

\abstract{Precise knowledge of the cross sections for neutrino interactions with nuclei is important not only for existing experiments but also for development of future detectors. When momentum transferred to the nucleus by the probe is high enough, the impulse approximation is valid, i.e. the nucleus can be described as composed of independent nucleons and the interaction happens between neutrino and a~single bound nucleon. The cross section of the nucleus is then expressed as an integral of the free cross section (modified by the off-shell effect) folded with the so-called spectral function, describing the distribution of nucleon momenta and energies.

I present results obtained for the approximate spectral functions of argon and calcium. The accuracy of the approach is verified by comparison with a~broad spectrum of precise electron-scattering data for the calcium target. Some of these data lie in the kinematical region corresponding to neutrino interactions what indirectly shows that a~similar level of accuracy is achieved for neutrinos. The calculated $\isotope{Ar}(\nu_\mu,\mu^-)$ cross section is significantly lower than the one predicted within the Fermi gas model.

}

\FullConference{10th International Workshop on Neutrino Factories, Super beams and Beta beams\\
		 June 30 - July 5 2008\\
		 Valencia, Spain}

\begin{document}
In the impulse approximation (IA) approach, lepton scattering off the nucleus is described as an interaction between the lepton and a~single nucleon, because the nucleus is treated as a~collection of independent nucleons. Then, to characterize the nucleus, one needs to know the distribution of momenta and energies of the nucleons that compose it. This distribution is called the spectral function. For the IA to be justified, the momentum transferred by the lepton,~$\n q$, must not be too low since the penetrated region of the nucleus, of order $\sim$$1/\n q$, have to cover only one nucleon.

To improve the description of neutrino-argon interactions in the sub-GeV energy range, I developed a~method to approximate spectral functions for medium-mass nuclei~\cite{ref:Ankowski}. All the details and the parametrization of the spectral functions of argon and calcium are given in Ref.~\cite{ref:Ankowski&Sobczyk}.

In quasielastic electron scattering, the only uncertain element is the description of the nucleus. Therefore I restrict myself to the quasielastic case in order to perform clear test of the obtained spectral functions: for low values of energy loss $\omega$, an agreement with experimental data should be obtained, and for high $\omega$'s, some cross section should be missing due to $\Delta$ excitation, nonresonant pion production, etc.

\begin{figure}
    \centering
    \includegraphics[width=1.0\textwidth]{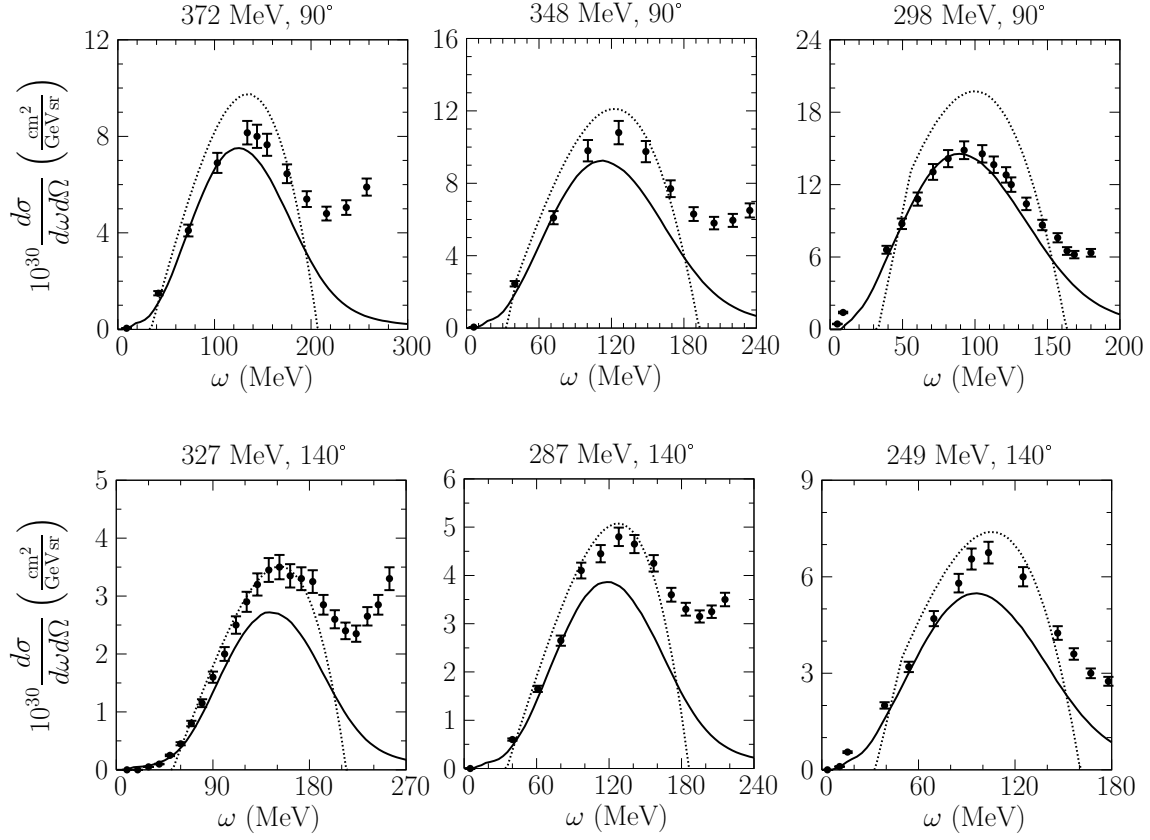}%
\caption{\label{fig:Ca} Quasielastic contribution to the $\isotope[40]{Ca}(e,e')$ cross section at various beam energies and scattering angles  calculated using the spectral function of calcium (solid line) and the Fermi gas model (dotted line). The experimental points for the inclusive cross section are from Ref.~\cite{ref:Williamson}.}
\end{figure}

Comparison of the calculated $\isotope[40][20]{Ca}(e,e')$ cross sections with the precise experimental data of Ref.~\cite{ref:Williamson}, collected at beam energies 130--841 MeV for scattering angles 45.5--140$^\circ$, showed very good accuracy of the approach, see the sample of the results in Fig.~\ref{fig:Ca} and in Ref.~\cite{ref:Ankowski&Sobczyk}. However, when the typical transferred momentum in the quasielastic peak decreases below $\sim$400 MeV/$c$, the IA starts to fail because the interaction region is big enough to cover two or more nucleons.

Since the electron data for scattering angle 45.5$^\circ$ kinematically correspond to neutrino interactions at energy similar to electron-beam energy~\cite{ref:Ankowski&Sobczyk}, one can expect that at energy of a~few hundreds of MeV nuclear effects in neutrino interactions are modeled with the accuracy similar to the case of electron scattering, wherever the IA is valid. The situation in neutrino physics is complicated by the fact that important observables include contribution of low-$\n q$ events, see more in Ref.~\cite{ref:Ankowski_IA}.

\begin{figure}
    \centering
    \includegraphics[width=0.65\textwidth]{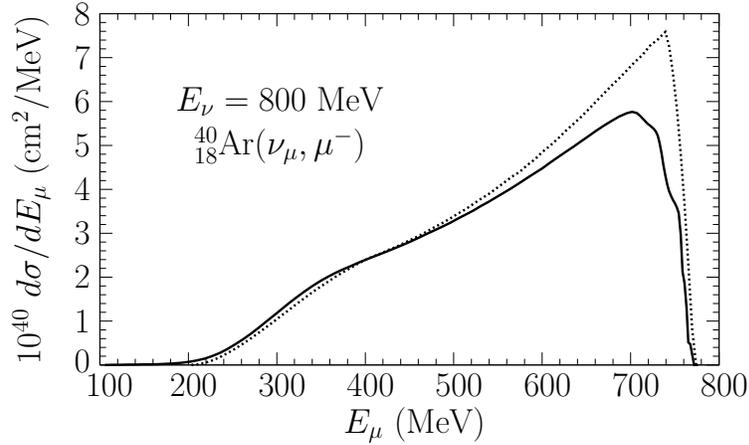}%
\caption{\label{fig:Ar} Quasielastic differential cross section $d\sigma/dE_\mu$ of the $\isotope[40][18]{Ar}(\nu_\mu,\mu^-)$ scattering as a~function of produced muon energy $E_\mu$ for the spectral function of argon (solid) and the Fermi gas model (dotted line).}
\end{figure}

Figure~\ref{fig:Ar} shows comparison of the differential $\isotope[40][18]{Ar}(\nu_\mu,\mu^-)$ cross section in produced muon energy $d\sigma/dE_\mu$ obtained using the spectral function from Ref.~\cite{ref:Ankowski&Sobczyk} and the Fermi gas model. Clearly visible difference between them, appearing for high values of $E_\mu$, is responsible for the total cross section to be 10.9\% lower in the case of the spectral function. A~correct description of the low-$\n q$ contribution (i.e. beyond the IA) would result in further reduction of the cross section in this region, increasing discrepancy between the Fermi gas model and the refined treatment of nuclear effects.

%When the IA is justified, nuclear effects in $\nu_\mu$ are accurately modeled by the presented approach.

\acknowledgments
I thank Jan T. Sobczyk for his contribution to this work and Giampaolo Co' for the calcium momentum distributions. This work was supported by MNiSW under by MNiSW under grant nos. 3735/H03/2006/31 and 3951/B/H03/2007/33.

%%%%%%%%%%%%%%%
%%%%%%%%%%%%%%%

%%%%%%%%%%%%%%%


\begin{thebibliography}{9}
    \bibitem{ref:Ankowski}%
        {A.~M.~Ankowski, Acta Phys. Pol.~B {\bf37}, 2259 (2006).}%
        %
    \bibitem{ref:Ankowski&Sobczyk}%
        {A.~M.~Ankowski and J.~T.~Sobczyk, Phys. Rev. C {\bf 77}, 044311 (2008).}%
        %
    \bibitem{ref:Williamson}%
        {C.~F.~Williamson {\it et al.}, Phys. Rev. C {\bf 56}, 3152 (1997).}%
        %
    \bibitem{ref:Ankowski_IA}%
        {A.~M.~Ankowski, PoS(Nufact08)118.}%
        %
\end{thebibliography}
\end{document}